\documentclass{pasj00}
\draft

\begin{document}
\SetRunningHead{S. Katsuda and H. Tsunemi}{Spectral Analysis of Vela Shrapnel D}
\Received{2005/1/4} 
\Accepted{2005/6/13} 

\title{Spatially Resolved Spectral Analysis of Vela Shrapnel D}

\author{Satoru \textsc{Katsuda} and 
Hiroshi \textsc{Tsunemi}
}
\affil{Department of Earth and Space Science,
	Graduate School of Science, Osaka University,\\
	1-1 Machikaneyama, Toyonaka, Osaka 560-0043}
\email{katsuda@ess.sci.osaka-u.ac.jp, tsunemi@ess.sci.osaka-u.ac.jp,\\}


\maketitle

\begin{abstract}

    The ROSAT all-sky survey discovered several `shrapnels', showing boomerang structures outside the Vela supernova remnant.  We observed shrapnel D with the XMM-Newton satellite.  There is an X-ray bright ridge structure in our FOV running from north to south.  Applying the VNEI model to X-ray spectra of various regions, we find that the plasma in the eastern part from the X-ray ridge is significantly different from that in the western part.  The X-ray spectra in the western part can be represented by a single-temperature component.  The abundances of heavy elements are almost uniform, whereas they are heavily overabundant, except for Fe; the relative abundances to the solar values are O $\sim5$, Ne $\sim10$, Mg $\sim10$, Fe $\sim1$.  This indicates that shrapnel D originated from the ejecta of the supernova.  We find that the plasma in the eastern part from the ridge consists of two components with different temperatures; the hot component comes from the ejecta, while the cold component comes from the interstellar matter.  These two components are considered to be in contact with each other, forming a contact discontinuity.  Around the northern part of the contact discontinuity, we find wave-like structures of which the typical scale are comparable with that of the Rayleigh--Taylor instability.
 
\end{abstract}

\section{Introduction}
 
 The Vela supernova remnant (SNR) is a middle-aged SNR that exploded about 10000 years ago (Taylor et al.\ 1993).  It clearly shows a circular structure with a diameter of 8$^\circ$.3 (Aschenbach et al.\ 1995).  At a distance of 250 $\pm$ 30 pc (Cha et al.\ 1999), this corresponds to 36 pc.  The Vela SNR is so close to us that it can be an ideal candidate for resolving fine structure in X-rays.  Aschenbach et al.\ (1995) also discovered `shrapnels', boomerang structures from `A' to `F', outside the main shell.  The opening angles of the shrapnels suggest supersonic motion in tenuous matter.  The center line of each boomerang structure can be traced back to the center of the SNR, which is close to the Vela pulsar.  Therefore, they are considered to originate from fragments of supernova (SN) ejecta that are beyond the position of the main blast wave.  There is an alternative explanation that the features are ``break-outs" of shock in which inhomogeneities in the ambient medium cause the shock to be non-spherical. 

 Vela shrapnel A was observed with ASCA (Tsunemi et al.\ 1999).  It was clarified that the abundance of Si is about 10-times higher than that of O.  Therefore, they concluded that shrapnel A was an explosion ejecta from a Si-rich layer of a progenitor star.  If it is ejecta of a SN explosion, the interstellar matter would be swept up in the leading edge while the ejecta material would be peeled off in the trailing edge.  However, the spatial distribution of Si could not be measured due to the poor spatial-resolving power of ASCA.  In order to investigate the spatial structure of shrapnel A, Miyata et al.\ (2001) observed it with the Chandra X-ray Observatory.  The Chandra image reveals a bright X-ray region at the head position of shrapnel A and a fainter extended tail.  They confirmed an overabundance of Si in the head region, but could not obtain a strong constraint on the parameters of the tail region.

 Shrapnel D, the eastern limb of the Vela SNR, is the closest to the main shell and the brightest in X-rays of the six shrapnels.  The optical nebula RCW 37 (Rodgers et al.\ 1960) lies along the outer edge of shrapnel D, which clearly shows that shrapnel D is now interacting with an interstellar cloud.  Sankrit et al.\ (2003) observed this optical filament and found that shrapnel D is a bow shock propagating into an interstellar cloud with normal abundances.  In order to understand the nature of shrapnel D, Plucinsky et al.\ (2002) observed it two times with Chandra, one at the head of shrapnel and the other in the trailing ``wake''.  If the X-ray emission associated with the shrapnel D is produced by shock-heating of the ambient medium by supersonic motion of the ejecta, one would expect the abundances to be enhanced at the head region and close to the interstellar medium (ISM) value at the wake region.  They found that the spectra from the different locations in the shrapnel are remarkably similar to each other.  They did not show the abundance variations that might be expected from a fragment of ejecta.  However, they could not obtain strong constraints on the elemental abundances: the O and Ne abundances vary from 0.5 to 5.0 $\times$ solar and 1.6 to 6.4 $\times$ solar, respectively.  They therefore concluded that the origin of shrapnel D is more consistent with a shock breakout hypothesis than with an ejecta hypothesis, although they cannot rule out the ejecta hypothesis. 

 What is indeed the origin of shrapnel D?  If it is a shrapnel of ejecta, how is the distribution of the element abundances in it?  In order to answer these questions, we observed shrapnel D with the XMM-Newton satellite.  We here present the results from the observation.  Then, we performed a spatially resolved spectral analysis, and showed the results of an in-situ observation of interaction zones between shrapnel D and the interstellar cloud.

\section{Observations}

\begin{figure}
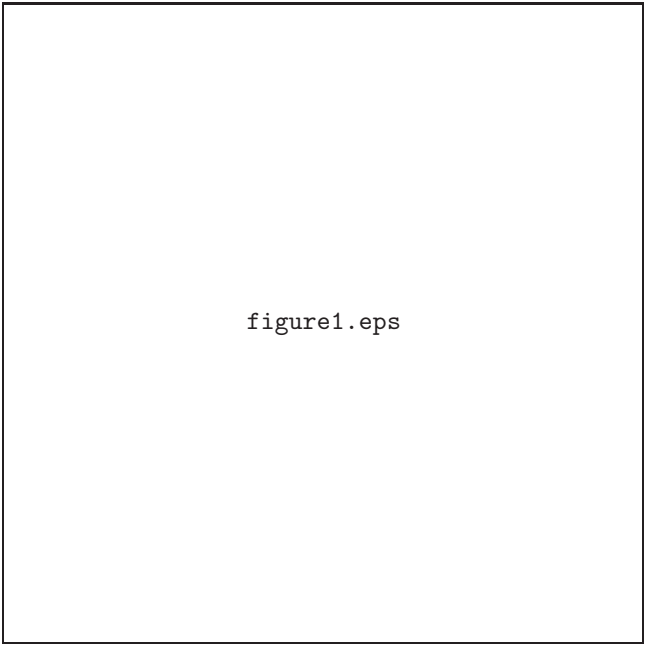

\FigureFile(85mm,85mm){figure1.eps}
\caption{ROSAT PSPC logarithmically scaled image of the Eastern Limb of the Vela SNR.  The black circle indicates the FOV of our XMM-Newton observation.}
 \label{FOV} 
\end{figure}

 We observed shrapnel D on 2002 May 2 with the EPIC (European Photon Imaging Camera) instruments on board the XMM-Newton satellite.  The observation ID is 0136010101 (PI H. Tsunemi).  Since the shrapnel is larger than the field of view (FOV) of the EPIC ($30^{\prime}$), we selected the brightest region in the shrapnel.  Figure \ref{FOV} shows a ROSAT PSPC image of shrapnel D with an overlay of the FOV of our observation.  The EPIC MOS cameras were operated in the standard full-frame mode and the EPIC PN camera in the extended full-frame mode.  Each camera was operated using a medium filter.  Since we screened the data by rejecting the high-background periods, the effective exposure time was reduced to $\sim$ 21 ks for MOS-1 and MOS-2 and to $\sim$ 14 ks for PN.  We selected X-ray events corresponding to patterns 0--12 for MOS and patterns 0--4 for PN, respectively.

 Since shrapnel D is larger than the FOV, we could not subtract the background events from the same data. Therefore, we used an observation (ID 0147511701) of the Lockman hole as background data.           

\section{Image Analysis}

\begin{figure}
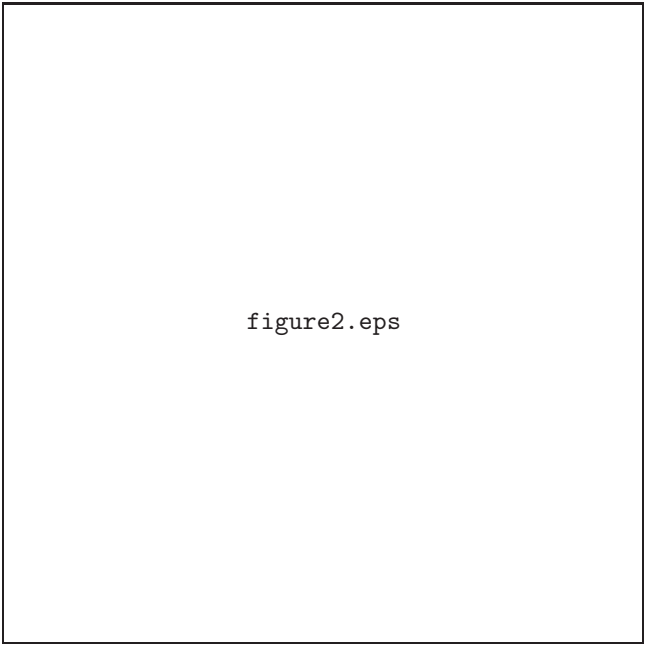

\FigureFile(85mm,85mm){figure2.eps}
\caption{EPIC MOS-1\,$+$\,MOS-2 logarithmically scaled image in the energy range 0.3--2.0\,keV.  The data have been smoothed with a Gaussian of $\sigma = 6''$.  The contours show a linearly scaled optical intensity map.}
 \label{Xim_opt} 
\end{figure}

 Figure \ref{Xim_opt} shows an exposure-corrected MOS-1\,$+$\,MOS-2 image of shrapnel D in the energy range of 0.3--2\,keV with an overlaid optical contour map.  We extracted the exposure map using SAS v\,5.4.1(the EEXPMAP command), and smoothed the image by a Gaussian of $\sigma$ = $6''$.  We found a bright X-ray ridge structure running from north to south in the region east of the FOV.  The X-ray emission from the eastern part of the ridge abruptly weakens to the background level, while that from the western part of the X-ray ridge gradually weakens.  There is also an optical bright ridge structure, a pencil nebula RCW 37, running parallel to the X-ray ridge at about $3^\prime$ in the east.  This suggests that shrapnel D is now interacting with an interstellar cloud (Sankrit et al.\ 2003).   

\section{Spectral Analysis}

   We investigated spectral variations from various regions in detail.  According to the X-ray ridge structure, we divided our FOV into many small rectangles ($2^\prime\times4^\prime$, a side with $2^\prime$ is perpendicular to the ridge and that of $4^\prime$ is parallel to that).  The scale of the rectangle is larger than the half-power beam width of XMM-Newton.  The scale was selected such that we could obtain sufficient photon statistics in each rectangle.  For the purpose of more detailed investigations, we arranged rectangles such that each rectangle overlapped each other by half of its size, obtaining 243 spectra in total.  Figure \ref{region} shows those regions overlaid on a MOS-1\,$+$\,MOS-2 image that is the same as figure \ref{Xim_opt}.  In the spectral analysis, we included the uncertainties among CCDs by 5\% on the model in quadrature (see, e.g., Nevalainen et al.\ 2003; Kirsch 2004).  We found that the spectra from the western part of the X-ray ridge were quite different from those from the eastern part (the boundary is indicated in figure \ref{region}).  Therefore, we divided the data into two regions by the X-ray ridge.  We present our results in the following subsections.

\subsection{Western Part of the X-Ray Ridge}

   Figure \ref{specA} shows an example spectrum from the white rectangle in figure \ref{region}.  We can see prominent O {\scshape VII} triplets at $\sim0.57$\,keV, O {\scshape VIII} Ly$\alpha$ at $\sim0.65$\,keV, Ne {\scshape IX} triplets at $\sim0.91$\,keV, Ne X Ly$\alpha$ at $\sim1.02$\,keV, Mg {\scshape XI} triplets at $\sim1.35$\,keV lines. These features are clearly seen in all of the spectra from the western part of the X-ray ridge.  We applied an absorbed non equiliblium ionization model (the VNEI model in XSPEC v\,11.3.1) to each spectrum (Hamilton et al.\ 1983; Borkowski et al.\ 1994, 2001; Liedahl et al.\ 1995).  The free parameters in our analysis are the electron temperature, $kT_\mathrm{e}$; the ionization parameter, $\tau$; the emission measure (hereafter EM; EM = $\int n_\mathrm{e}n_\mathrm{i} dl$, where $n_\mathrm{e}$ and $n_\mathrm{i}$ are the number densities of electrons and ions, respectively, and $dl$ is the plasma depth); the abundances and the column density, $N_\mathrm{H}$, of the absorbing foreground material.  Above, $\tau$ is the electron density times the elapsed time after shock heating.  Only the abundances of C, N, O, Ne, Mg, Fe are free, and the others are frozen to the solar values (Anders, Grevesse 1989).  We set the abundances of C and N equal to that of O.  All of the spectra are represented by a single-temperature VNEI model. 

The reduced $\chi^2$ is 1--1.5 for about 85\% of the region to the west of the X-ray ridge.  The remaining (only 15\% in the west of the ridge) regions, which are located near the ridge and have a high surface brightness, show about 1.5--1.8.  These values are far from acceptable from a statistical point of view.  These regions are very close to the interaction region with the interstellar cloud.  We could not obtain better fits by adding extra components there.  Therefore, our model is too simple to reproduce the entire region.  A more advanced model will be needed to obtain a better fit, particularly near the X-ray ridge.  In table 1, we show the best-fit parameters of the spectrum from the white rectangle.   

\begin{figure}
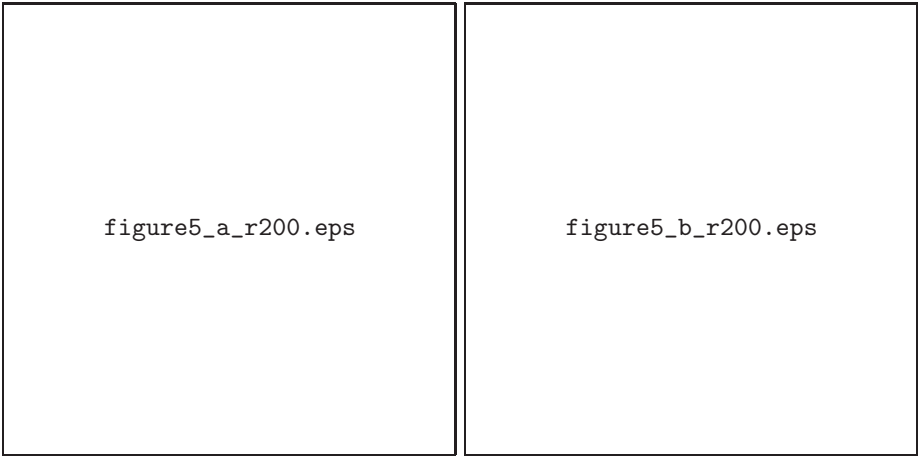

\FigureFile(60mm,60mm){figure5_a_r200.eps}
\FigureFile(60mm,60mm){figure5_b_r200.eps}

\caption{Linearly scaled contour map of $kT_\mathrm{e}$ and log\,$\tau$ in the western part of the X-ray ridge overlaid on an X-ray image the same as figure \ref{Xim_opt}.  The values of $kT_{\mathrm e}$ are in units of keV.  The data were smoothed with a Gaussian of $\sigma = 4'$. }
 \label{kTe_tau} 
\end{figure}

\begin{figure}
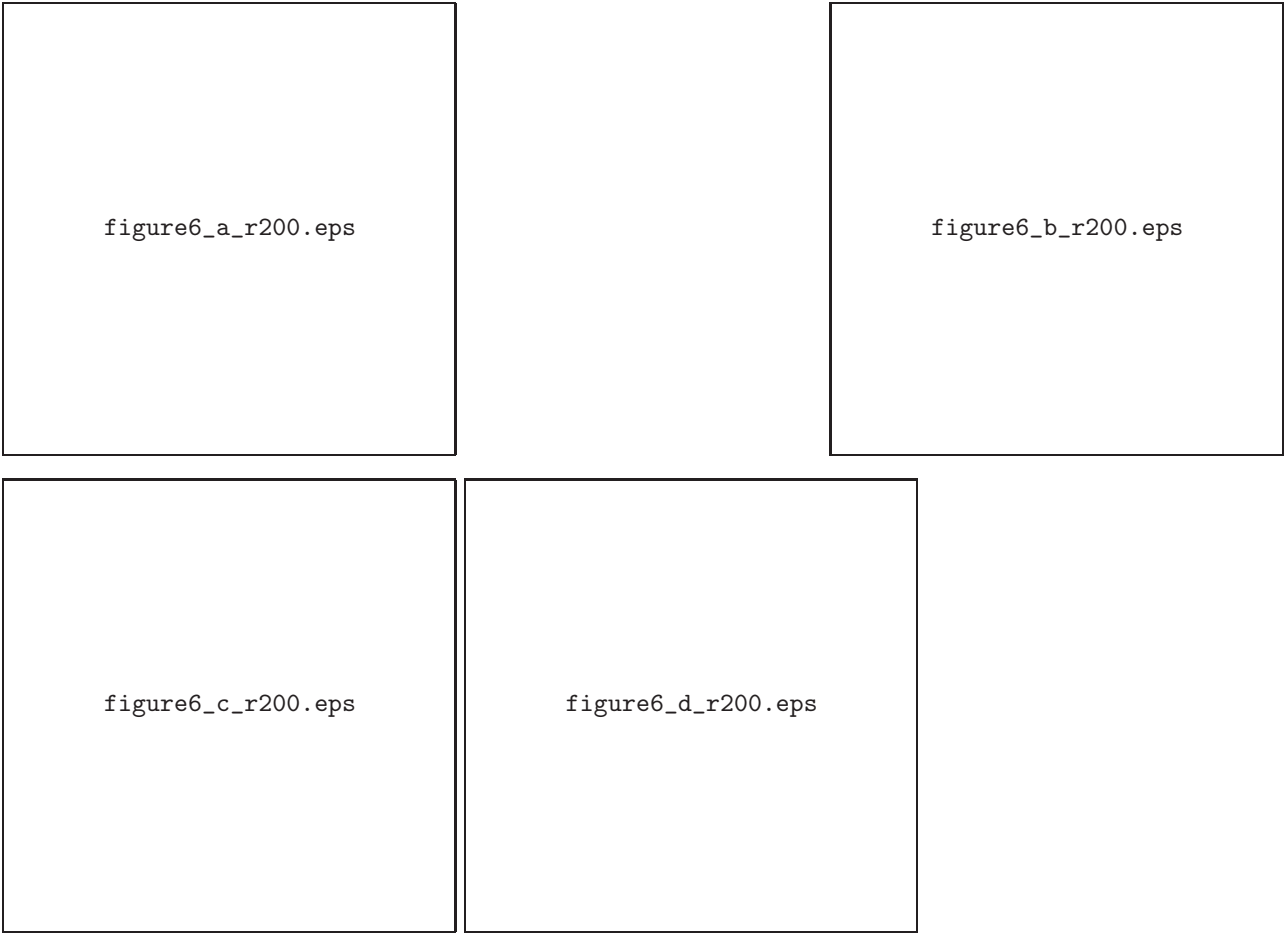

\FigureFile(60mm,60mm){figure6_a_r200.eps}
\FigureFile(60mm,60mm){figure6_b_r200.eps}
\FigureFile(60mm,60mm){figure6_c_r200.eps}
\FigureFile(60mm,60mm){figure6_d_r200.eps}

\caption{Linearly scaled contour map of each metal abundance overlaid on an X-ray image the same as figure \ref{Xim_opt}.  The values shown in each figure are those relative to the solar abundance.  The data were smoothed with a Gaussian of $\sigma = 4'$.}
 \label{abundances} 
\end{figure}

Figure \ref{kTe_tau} shows the values of $kT_\mathrm{e}$ and a log\,$\tau$ contour map in the western part of the X-ray ridge overlaid on the MOS-1\,$+$\,MOS-2 image.  The value of $kT_\mathrm{e}$ shows the highest (0.32\,$\pm0.01$\,keV) at the ridge and a gradual decrease in the west down to 0.25\,$\pm0.02$\,keV.  The spatial variation of log\,$\tau$ shows an anti-correlation with that of $kT_\mathrm{e}$.  Most of the regions, having relatively low $kT_\mathrm{e}$, show $\tau$ $>10^{12}\mathrm{s\,cm^{-3}}$, which indicates a collisional ionization equilibrium condition.  However, the regions near the X-ray ridge, having a relatively high $kT_\mathrm{e}$, show that $\tau$ is (4.8--7.2)$\times10^{11}\mathrm{s\,cm^{-3}}$, which indicates a nonequilibrium ionization condition.  Figure \ref{abundances} shows contour maps of metal abundances.  They are heavily overabundant, except for Fe, at any regions in the map; the values of the abundances relative to the solar values are O $\sim5$, Ne $\sim10$, Mg $\sim10$, Fe$\sim1$.  These values support the idea that shrapnel D originates from the explosion ejecta.  

\subsection{Eastern Part of the X-Ray Ridge}

   A weak X-ray emission comes from the eastern part of the X-ray ridge.  We found that the spectra from the eastern part of the X-ray ridge were quite different from those of the western part.  We divided the eastern part into two regions.  

\subsubsection{Southeastern region near the optical filament}

   We found that the spectrum from the region near the optical filament could be well fitted with a single component with solar abundances.  It shows a relatively low $kT_\mathrm{e}$ ($\sim$\,0.15\,keV).  Sankrit et al.\ (2003) found that optical emission comes from a shock-excited cloud with solar abundances.  Therefore, we conclude that the low-$kT_\mathrm{e}$ component represents plasma originating from the cloud.  

 We also found that the spectrum varies between the optical filament and the X-ray ridge.  We analyzed the spatial variation of the spectra between them.  We extracted spectra from rectangular regions ($1^\prime\times8^\prime$) drawn by the white line in figure \ref{reg_Omap} a, each of which overlapped adjacent regions by the half width. In this way we extracted 5 spectra.  We found that the spectra were not represented by a single-temperature VNEI model.  Therefore, we added an extra component with different $kT_\mathrm{e}$.  Due to the insufficient photon statistics of the low-$kT_\mathrm{e}$ component above the O-line energy, we assumed solar abundances for all elements, except O.  The extra component improves the fit (e.g., reduced $\chi^2$ is improved from 1.7 to 1.2), which is in contrast to that of the western part.  The $kT_\mathrm{e}$ values of these two components are 0.15\,keV and 0.35\,keV, while the intensity fraction varies from east to west.  The abundances of the high $kT_\mathrm{e}$ component are similar to those of the western part of the X-ray ridge, which shows that it comes from the ejecta.  The abundance of O and $kT_\mathrm{e}$ for the low-$kT_\mathrm{e}$ component are similar to those for the cloud component.  Therefore, the low-$kT_\mathrm{e}$ component is considered to originate from the cloud.  The fraction of the cloud component decreases toward the X-ray ridge.  We show the variation of EM corrected by the interstellar absorption as a function of the location in figure \ref{EMvariation}.  The EM of the cloud component decreases toward the X-ray ridge, while that of the ejecta increases toward the same direction.  It is reasonable to consider that this feature is due to the projection effect, as looking obliquely between the cloud and the ejecta.  We consider that the boundary region between these two plasmas is a contact discontinuity.

\subsubsection{Northeastern region} 

  In the X-ray surface brightness map, we found wavelike structures at the northeastern region of the FOV in figure \ref{reg_Omap} a.  Figure \ref{reg_Omap} b shows a band ratio map of the O\,{\scshape VIII} band (624--684\,eV) / the O\,{\scshape VII} band (545--603\,eV), in which the wavelike structures are prominent.  The arrow in figure \ref{reg_Omap} b shows $\lambda_\mathrm{o}$, a typical length of the wavelike structure, to be $\sim0.3$ pc.  

 In order to investigate the feature of the wavelike structure, we extracted spectra from regions indicated in figure \ref{reg_Omap} b.  The circular regions with the thick solid line and dashed line are responsible for the regions where the O\,{\scshape VIII} band is dominant, while those with thin line are responsible for the regions where the O\,{\scshape VII} band is dominant.  All of the spectra are represented by a two-component VNEI model, just as those applied in the southeastern region.  It is reasonable to consider that the low-temperature component represents an interstellar cloud, while the high-temperature component represents the ejecta.  Then, the cloud dominates in the circular regions with thin line and the ejecta dominate in the circular regions with thick line.  We consider that there is a contact discontinuity in those regions.  Figure \ref{spec12} shows the spectrum from the dashed circle indicated in figure \ref{reg_Omap} b and the best-fit parameters are given in table 1.  The value of the plasma depth of the ejecta comes from the assumption that the circular shape in figure \ref{reg_Omap} is due to the spherical shape.  This result suggests that the surface of the contact discontinuity has wavelike structures with a typical length, $\lambda_\mathrm{o}$. 

\section{Discussion}

  X-ray spectra in the western region of the X-ray ridge can be represented with a single-temperature VNEI model.  The abundances of heavy elements, except for Fe, are significantly higher than those of solar values in any region in the western part of the X-ray ridge: O $\sim5$, Ne $\sim10$, Mg $\sim10$, Fe $\sim1$.  The absolute abundance obtained by the spectral fits depends on the model employed, whereas the relative abundances among the heavy elements are robust.  The relative abundances, (O/Fe, Ne/Fe and Mg/Fe) are $\sim5$, $\sim10$ and $\sim10$, respectively.  High-Z elements, like Ar, Ca, and Fe, are generated dominantly in a Type-{\scshape I}a SN (Nomoto et al.\ 1984), while low-Z elements, like O, Ne, Mg, and Si, are generated in a Type-{\scshape II} SN (Thielemann et al.\ 1996).  The relative abundances indicate a Type-{\scshape II} SN origin rather than Type-{\scshape I}a.  Since Vela SNR is believed to be a Type-{\scshape II} SN, we conclude that shrapnel D is ejecta originating from an explosion of Vela SN.

 We estimated the total mass of the ejecta of Vela shrapnel D.   We assumed that shrapnel D has a conical structure with a diameter 80$^\prime$ and a height of 45$^\prime$.  Using a distance of 250 pc, and the assumption of uniform electron density of $\sim\mathrm{0.2\,cm^{-3}}$, we can estimate the total mass to be about $10^{-1} \,\mathrm{M_{\odot}}$. 

 We found that shrapnel D is ejecta coming from O, Ne and Mg-rich layer of a progenitor star.  This is contrast to the fact that shrapnel A comes from the Si rich layer (Tsunemi et al.\ 1999, Miyata et al.\ 2001).  These findings suggest that shrapnel D comes from an outer layer of the progenitor star than shrapnel A.  We compare the present distance from the center of Vela SNR to shrapnel D with that to shrapnel A.  Aschenbach et al.\ (1995) inferred the temperature of shrapnel D to be 0.34$\pm$0.07\,keV from the opening angle of the feature and the temperature of the ambient medium.  The opening angle was estimated based on the assumption that the shrapnel is moving in the plane of the sky.  The measured temperature (0.32$\pm$0.01\,keV) is in good agreement with that inferred from the geometry, which supports that shrapnel D is moving in the plane of the sky.  Therefore, the actual distance of shrapnel D from the center of Vela SNR is shorter than that of shrapnel A.  Since shrapnel D is considered to have come from the outer layer of a progenitor star than shrapnel A, it suggests that a part of the inner layer has a higher initial velocity than that of the outer layer.

 In the southeastern region of the FOV, there is an optical filament, a pencil nebula, RCW 37, running parallel to the X-ray ridge about $3^{\prime}$ to the east.  We confirmed that the emission from the optical filament region comes from a shock-excited cloud (Sankrit et al.\ 2003).  We found the spectrum variation in the region between the X-ray ridge and the pencil nebula.  An extra component is needed to fit the spectrum.  The low-temperature component shows solar abundances, while the high-temperature component shows high metal abundances.  The fraction of the low-temperature component gradually decreases toward the X-ray ridge, while that of the high-temperature component increases.  Therefore, the low-temperature component is considered to come from the cloud and the high-temperature component is considered to come from the ejecta.  

 We found wavelike structures in the northeastern region of the FOV in the X-ray surface brightness map.  This structure is more apparent in the band ratio map between the O\,{\scshape VIII} band and the O\,{\scshape VII} band.  Furthermore, our spectral analysis indicates that a contact discontinuity between the ejecta and the interstellar cloud has wavelike structures.  The wavelike structures remind us of the Rayleigh--Taylor (R--T) instability, which occurs where a heavy fluid is accelerated by a light one (i.e., the pressure of the light fluid is higher than that of the heavy one).  In our case, the temperature of the measured ejecta ($T_\mathrm{ej}$) exceeds that of the cloud ($T_\mathrm{cl}$).  Therefore, if this is the case, the heavy fluid must be the cloud and the light one must be the ejecta.  The typical scale of R--T instability is given by (Velazquez et al.\ 1998)

\[\lambda = 0.39\,\times\left( \frac{T_{7}^5}{g_{2}\,\alpha\,n_{1}^2} \right)^{1/3} \mathrm{pc}, \]
where 
\[T_{7}=\frac{T_\mathrm{cl}}{10^7\,\mathrm{K}},\,\,n_\mathrm{1}=\frac{n_\mathrm{cl}}{1\,\mathrm{cm^{-3}}},\]
\[g_\mathrm{2}=\frac{g_\mathrm{eff}}{10^{-2}\,\mathrm{cm\,s^{-2}}},\,\,\alpha = \left(\frac{n_\mathrm{cl} - n_\mathrm{ej}}{n_\mathrm{cl} + n_\mathrm{ej}} \right). \]

Here, $g_\mathrm{eff}$ is the effective gravity (inertial force on a frame of the contact discontinuity), $n_\mathrm{cl}$, the number density of the cloud and $n_\mathrm{ej}$ that of the ejecta.  We assumed the cloud depth to be 0.73 pc (= 10$^\prime$ ) in order to estimate $n_\mathrm{cl}$.  The value of $g_\mathrm{eff}$ is given by ($1/\rho_\mathrm{m}$)($dP/d\mathrm{r}$) (Ebisuzaki et al.\ 1989), where $dP/d\mathrm{r}$ is the pressure gradient and $\rho_\mathrm{m}$ is the mean density between the cloud and the ejecta.  We used the approximation that $dP/d\mathrm{r} \sim \Delta P/{\lambda}$, where $\Delta P$ represents the pressure difference between two fluids.  We consider that the temperatures of the cloud and the ejecta come from shock heating.  After shock heating, the ion temperature would not be equal to that of the electron.  Then, the electron and ion temperatures vary smoothly to reach the thermal equilibrium.  The postshock electron and ion temperatures equilibrate rapidly after a slow shock, even when the shock is collisionless (Ghavamian et al.\ 2001).  Therefore, we assumed thermal equilibrium between the ion and the electron, $P_\mathrm{i} = P_\mathrm{e}$ ($P_\mathrm{i} , P_\mathrm{e}$ are the ion and the electron pressure, respectively), because the shock speed is only about 150 $\mathrm{km\,s^{-1}}$ (Sankrit et al.\ 2003) in our case.  Based on the assumption of $P_\mathrm{i} = P_\mathrm{e}$, the pressure difference between the cloud and the ejecta becomes $2\times\Delta P_\mathrm{e}$ ($\Delta P_\mathrm{e}$ is the electron pressure difference between the cloud and the ejecta).  Using the values obtained by the spectral fitting (from the dashed circle shown in table 1), we obtained $\lambda$ as    

\[\lambda \sim 0.2\,\times \left(\frac{T_\mathrm{7}}{0.186}\right)^{5/2} \left(\frac{n_\mathrm{m1}}{0.7}\right)^{1/2}  \left(\frac{P_\mathrm{10}}{3.7}\right)^{-1/2}  \left(\frac{\alpha}{0.1}\right)^{-1/2}  \left(\frac{n_\mathrm{1}}{0.77}\right)^{-1} \mathrm{pc}, \]
where
\[n_\mathrm{m1} = \frac{\rho_\mathrm{m}}{1.67\,\times10^{-24}\,\mathrm{g\,cm^{-3}}}, \]
\[P_\mathrm{10} = \frac{\Delta P}{1\,\times\,10^{-10}\,\mathrm{erg\,cm^{-3}}}.\]

 We found that $\lambda$ is close to $\lambda_\mathrm{o}$.  Since the value of $n$ strongly depends on the plasma depth, the value of $\lambda$ also depends on the plasma depth.  It is highly probable that the wavelike structures are caused by the R--T instability.

  In the southeastern region of the FOV (around the optical filament), we cannot see wavelike structures.  In this region, we find $n_\mathrm{ej}\,<\,n_\mathrm{cl}$, and the pressure of the ejecta is lower than that of the cloud, where the contact discontinuity is stable.

\section{Conclusion}

 We observed shrapnel D with the XMM-Newton satellite.  We found an X-ray bright ridge structure running from north to south.  We found that the spectra from the western part of the X-ray ridge could be represented by a single $kT_\mathrm{e}$ component with high metal abundances: O $\sim5$, Ne $\sim10$, Mg $\sim10$, Fe $\sim1$.  This clearly shows that the origin of shrapnel D is ejecta from Vela SN.  The spectra from the eastern part of the X-ray ridge were represented by two $kT_\mathrm{e}$ components.  The low-temperature component having solar abundance is considered to come from the interstellar cloud, while the high-temperature component is considered to come from the ejecta.  This is the place where the ejecta are interacting with the interstellar cloud.  In the southeastern region of the FOV, it is reasonable to think that we see the interaction obliquely due to the projection effect.  In the northeastern region of the FOV, it is highly probable that we see the wavelike structure of the R--T instability. 

\bigskip

The authors would like to express their special thanks to Mr. H. Enoguchi. This work is partly supported by a Grant-in-Aid for Scientific Research by the Ministry of Education, Culture, Sports, Science and Technology (16002004).  This study is also carried out as part of the 21st Century COE Program, \lq{\it Towards a new basic science: depth and synthesis}\rq. H.T. is partly suppoted by the ISSI.

\newpage
\begin{table*}
\begin{center}
 \caption{Spectral-fit parameters.$^{\ast}$}
 \begin{tabular}{lcc}
  \hline\hline
 Parameter & White rectangle in figure \ref{region}  & Dashed circle in figure \ref{reg_Omap} \\
\hline
      $N_{\mathrm H} [10^{20}\,\mathrm{cm}^{-2}$] \dotfill & 2.9$^{+0.2}_{-0.4}$ &1.9$^{+0.2}_{-0.4}$ \\
   Low temperature component&&\\	
   $kT_\mathrm{e}$[keV] \dotfill & \dotfill&0.16$^{+0.01}_{-0.02}$  \\
   O(=C=N) \dotfill&\dotfill & 0.11$^{+0.06}_{-0.04}$  \\
   $\tau [\mathrm{s\,cm^{-3}}]$ \dotfill &\dotfill & $>3\times10^{11}$  \\ 
   EM$[\mathrm{cm^{-5}}]^{\dagger}$\dotfill&\dotfill&$(13\pm2)\times10^{17}$\\ 
   Plasma depth [pc]&\dotfill&0.73\\
   Electron density [$\mathrm{cm^{-3}}$]&\dotfill& 0.77$^{+0.07}_{-0.06}$ \\
   Electron pressure [$10^{-10} \mathrm{erg\,cm^{-3}}$]&\dotfill&1.99$\pm0.22$ \\
\hline
   High temperature component&&\\
$kT_{\mathrm e}$[keV] \dotfill &0.30$\pm$0.01   &0.38$\pm0.01$ \\
   O(=C=N) \dotfill &6.4$^{+0.2}_{-0.1}$& 4.5$^{+0.2}_{-0.1}$  \\
Ne \dotfill & 13.7$^{+0.8}_{-0.3}$ &11.5$^{+0.3}_{-0.8}$  \\
   Mg     \dotfill &14$\pm$2 &   8.9$^{+1.2}_{-2.1}$  \\
   Fe \dotfill & 1.4$^{+0.2}_{-0.1}$  & 1.00$^{+0.13}_{-0.15}$  \\
   $\tau [\mathrm{s\,cm^{-3}}]$ \dotfill&$(6.1\pm0.2)\times10^{11}$ & $(1.9\pm0.06)\times10^{11}$  \\
   EM$[\mathrm{cm^{-5}}]^{\dagger}$\dotfill&$(2.60\pm0.02)\times10^{17}$ & $2.66^{+0.05}_{-0.06}\times10^{17}$\\
   Plasma depth [pc]&2.2&0.2\\
   Electron density [$\mathrm{cm^{-3}}$]&0.197$\pm0.001$&0.627$\pm0.007$\\
   Electron pressure [$10^{-10} \mathrm{erg\,cm^{-3}}$]&0.96$\pm0.06$&3.86$\pm0.25$\\
   $\chi^2$/d.o.f. \dotfill &436/303 & 362/257 \\
   \hline
\\[-8pt]
  \multicolumn{3}{@{}l@{}}{\hbox to 0pt{\parbox{140mm}{\footnotesize
\par\noindent
\footnotemark[$*$]Other elements are fixed to those of solar values.  The values of abundances are multiples of solar value.  The errors are in the range $\Delta\,\chi^2\,<\,2.7$ on one parameter.  In the fits, 5\% systematic errors are included.
\par\noindent
\footnotemark[$\dagger$]EM denotes the emission measure $\int n_\mathrm{e} n_\mathrm{i} dl$.
}\hss}}

  \end{tabular}
 \end{center}
\end{table*}

\begin{figure}
\FigureFile(85mm,85mm){figure3.eps}
\caption{Same as figure \ref{Xim_opt} with an overlaid all the regions where we extracted spectra.  The spectrum from the white rectangle is shown in figure \ref{specA}.}
 \label{region} 
\end{figure}

\begin{figure}
\FigureFile(85mm,85mm){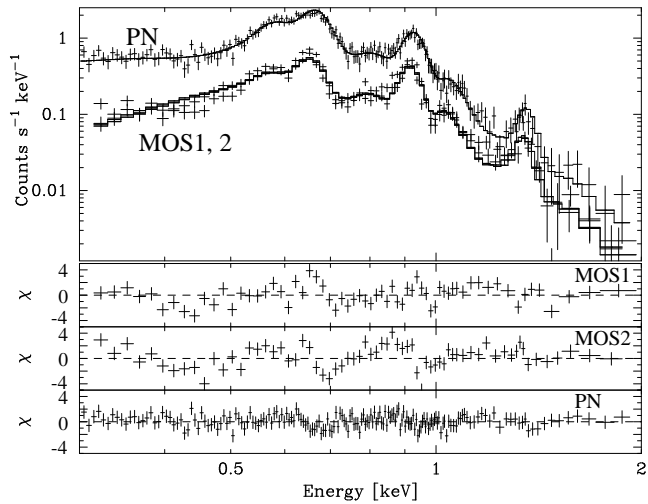}
\caption{X-ray spectra extracted from the white rectangle in figure \ref{region}.  The best-fit curves are shown with solid lines and the lower panels show the residuals.}
 \label{specA} 
\end{figure}

\begin{figure}
\FigureFile(85mm,85mm){figure7_a050710.eps}
\FigureFile(85mm,85mm){figure7_b050710.eps}
\caption{(a) Same as figure \ref{Xim_opt}. The white rectangles adjacently lined up are the regions where we extracted the spectra.  The large upper-left region indicates the region where we extracted the band ratio map.  (b) EPIC (MOS-1\,$+$\,MOS-2\,$+$\,PN) band ratio map (O\,{\scshape VIII} band / O\,{\scshape VII} band).  The circles are the regions where we extracted spectra.}
 \label{reg_Omap} 
\end{figure}

\begin{figure}
\FigureFile(85mm,85mm){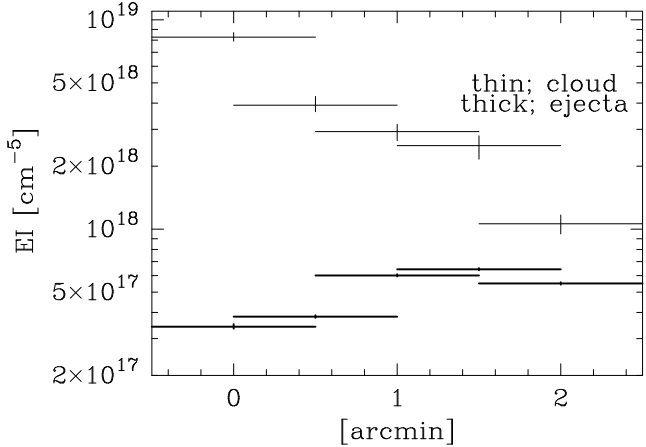}
\caption{Variation of EM of two components in the region between the optical filament and the X-ray ridge indicated in figure \ref{reg_Omap} by white rectangles.}
 \label{EMvariation} 
\end{figure}

\begin{figure}
\FigureFile(85mm,85mm){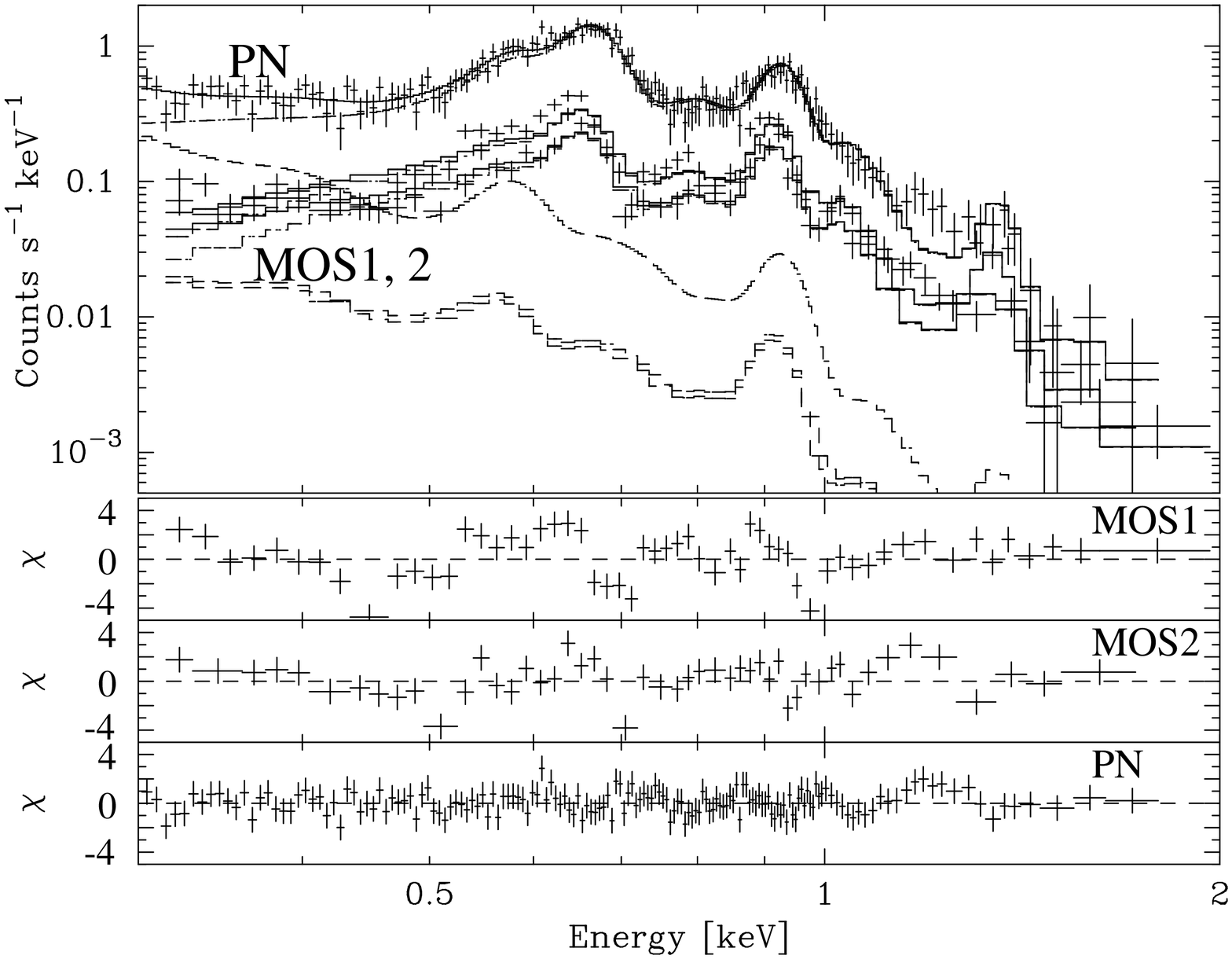}
\caption{X-ray spectra extracted from the dashed circle in figure \ref{reg_Omap} b.  The solid line is the total model, while the broken lines represent the individual contributions.  The lower panels show the residuals.}
 \label{spec12} 
\end{figure}

\end{document}